\documentclass[%
superscriptaddress,
amsmath,amssymb,
aps,
twocolumn,
]{revtex4-2}
\usepackage{hyperref}
\hypersetup{colorlinks,allcolors=blue}
\usepackage{graphicx}% Include figure files
\usepackage{bm}% bold math
\usepackage[all]{hypcap} % hyperref to top of figure
\usepackage{color}
\usepackage{siunitx}
\usepackage{mathtools}
\usepackage{ulem}
\newcommand{\q}[1]{``#1''}
\begin{document}

\title{Controllable spin filtering and half metallicity in $\beta_{12}$-borophene nanoribbons}

\author{F. Norouzi}
\affiliation{\footnotesize{Department of Physics, Iran University of Science and Technology, Narmak, Tehran 16844, Iran}}

\author{M. Farokhnezhad}
\affiliation{\footnotesize{School of Nano Science, Institute for Research in Fundamental Sciences (IPM), Tehran 19395-5531, Iran}}

\author{M. Esmaeilzadeh}
\email{mahdi@iust.ac.ir}
\affiliation{\footnotesize{Department of Physics, Iran University of Science and Technology, Narmak, Tehran 16844, Iran}}

\author{B. Szafran}
\affiliation{\leftline{\footnotesize{AGH University of Science and Technology, Faculty of Physics and Applied Computer Science, al. Mickiewicza 30, 30-059 Krak\'ow, Poland}}}

\date{\today}

\begin{abstract}
The experimental observation of the Dirac fermion states in $\beta_{12}$-borophene sheets and the discovery of their novel topological properties have made them a promising candidate for spintronic applications. Here, by combining non-equilibrium Green's function (NEGF) and  tight-binding (TB) approximation, we study the charge and spin transport properties through a $\beta_{12}$-borophene nanoribbon (BNR) with the different edge shapes. We show when a BNR exposed to a nonlocal exchange magnetic field, the spin filtering occurs for both spin-up and spin-down so that the spin direction of transmitted electrons could be controlled by adjusting the energy of incoming electrons with the help of an external backgate voltage. It is found that an armchair BNR (ABNR) in the simultaneous presence of a transverse electric field and a nonlocal exchange field indicates a half-metallic nature which is electrically controllable. Moreover, the influence of local exchange fields is evaluated by exposing the edges of ABNR to ferromagnetic strips with parallel and anti-parallel configurations. Our findings show that the edge manipulations in ABNRs lead to the emergence of a giant magnetoresistance and a perfect spin filter. Finally, we studied the effects of edge vacancies and Anderson disorder on the spin-dependent conductance of an ABNR and find that the perfect spin polarization is not destroyed in the presence of Anderson disorder and various single vacancies. Our results reveal the outstanding spin transport properties of ABNRs for future spintronic devices.
\end{abstract}


\maketitle
\section{Introduction}
Borophene, a monolayer of boron atoms with honeycomb-like lattice was synthesized on Ag (111) substrate by MBE (Molecular Beam Epitaxy) under ultrahigh vacuum \cite{bib1, bib2}. The experimental observations on $\beta_{12}$-borophene, which is one of the monolayer boron phases with the planar structure, shows that it has a weak interaction with Ag (111) substrate \cite{bib1}. Recently, researchers experimentally confirmed the existence of gapless Dirac cones in borophene so that reported results reveal its potential applications in spintronic nanodevices \cite{bib3, bib4}. 

The boron element in borophene with a highly delocalized covalent bond and one less electron than carbon accounts a paramount element \cite{bib5, bib6, bib7}. Also, boron is a light atom that leads to weak spin-orbit interaction (SOI), so it is classified in a different topological class compared to carbon \cite{bib8, bib9}. The existence of (nc-2e) delocalized multicenter bonds in boron due to resolve electron-deficient has made various boron-based phases \cite{bib10, bib11, bib12, bib13}. Indeed, each of its phases has a special mean ratio which is defined as the hexagonal holes (HHs) to the number of atoms in a triangular lattice so that it distinguishes their electronic properties from each other \cite{bib14, bib15}. Moreover, the bonding state of boron is divided into two parts consisting of in-plane and out-plane parts \cite{bib8}. Here, the in-plane part (contribution from s, $p _x $, $p _y$ orbitals) helps to form $\sigma$ bonding and anti-bonding, while the out-plane part (contribution from $p_z$ orbital) forms half-full of $\pi$ bonding and anti-bonding \cite{bib16}. 

As mentioned before, $\beta _{12}$-borophene, similar to graphene and silicene hosts Dirac cones where $p_z$ plays a role as an essential orbital \cite{bib17, bib18, bib19, bib20}. In addition, borophene makes effective contact with various 2D semiconductors so that it can reduce the contact resistance and improve the performance of future 2D transistors \cite{bib21}. This distinctive feature would make borophene a promising candidate for electronic devices \cite{bib22}. As a matter of fact, $\beta _{12}$-borophene has a strong electron-phonon coupling leading to high conductance in ballistic-regime \cite{bib8, bib12, bib23, bib24, bib25}. Also, reviewing the binding complexity and structural diversity, borophene will be an exceptional 2D material in spintronics.

The  lateral confinement of borophene nanoribbon provides an additional degree of control for the electron structure to be used in spintronic applications. In addition, it will reveal its novel physical properties such as size-dependent band gap, edge-dependent conductance and edge states. Zhong \textit{et al.,} synthesized single-atom-thick borophene nanoribbons by self-assembly and EVBPD (Electron Vapor Beam Physical Deposition) of boron on Ag (110) \cite{bib26}. After these successful processes, it is possible to introduce the borophene with amazing abilities as a new member of electronic and spintronic devices. 

With the synthesis of BNRs, some early theoretical works were performed on their electronic properties. For example, some researchers using Density-functional theory (DFT) studied the electronic properties and stability of BNRs \cite{bib27,bib28}. Also, Ezawa derived a Dirac theory for $\beta _{12}$-borophene and constructed three-band theories for triplet fermions. In addition, the edge states of BNRs and their edge-dependent properties were studied \cite{bib29}. Then, Meng \textit{et al.} showed the effect of nanoribbon width, edge direction and hydrogenation on the electronic properties of BNRs. There are a few theoretical studies for spintronic applications in BNRs \cite{bib30}. Using first principle calculations and NEGF method, the spin transport in zigzag borophene nanoribbons (ZBNRs) with the effects of edge passivation was studied and observed spin-dependent negative differential resistivity (NDR) \cite{bib31}. In the same way, Liu \textit{et al.}  investigated the quantum transport and the spin-filtering effect on ZBNRs with different widths. They could observe the spin filtering about $\%36$ at a certain bias voltage \cite{bib32}. 
\setlength{\parskip}{0.2cm}

In the present paper, we study charge and spin transport properties in an armchair $\beta _{12}$-borophene nanoribbon with different terminated edges in the presence of local and nonlocal exchange magnetic fields using tight-binding (TB) model combined with NEGF method. Our calculations show that the degeneracy of three-band touching points reported in Ref. \cite{bib29} is lifted for ABNRs, but the remaining bands play a crucial role on their electronic structure. 

To control spin-dependent properties of transmitted electrons, an external electric field is applied transversely to the ABNR. Our results show that perfect ($\%100$) spin filtering could take place in the presence of nonlocal exchange field and the spin direction of electrons can be easily controlled via changing the transverse electric field. In this case, ABNRs behave like a half-metal system. For local exchange field, we consider both parallel and anti-parallel configurations of exchange fields which are induced by depositing the ferromagnetic strips on both edges of the nanoribbon. It is shown that for parallel configuration spin polarization could occur and its spin direction can be tuned by changing the electron Fermi energy. However, there is no spin filtering for anti-parallel configuration and the bands for spin up and spin down are degenerate. Also, we show that the local exchange fields offer a control on the spin transport similar to the one found in the giant magnetoresistance effect. Moreover, we take into account the effects of nonmagnetic Anderson (short range) disorder and different single vacancies on spin-dependent conductance. Our findings show that the perfect spin filtering is not destroyed in the presence of Anderson disorder and vacancy defects. 

The layout of this paper is arranged as follows. In Sec. \ref{sec2}, we introduce our model, which is utilized to calculate the electronic structure and the quantum transport properties. In Sec. \ref{sec3}, we are being to reveal a mechanism that enables us to persuade a perfect spin filtering in the structure of BNRs. In the Sec. \ref{sec4}, a summary and conclusion, are given.

\section{THEORETICAL MODEL} \label{sec2}
\subsection{Tight-binding Hamiltonian of $\beta _{12}$-borophene}
\begin{figure}[t]
\centering
\includegraphics[scale=0.65]{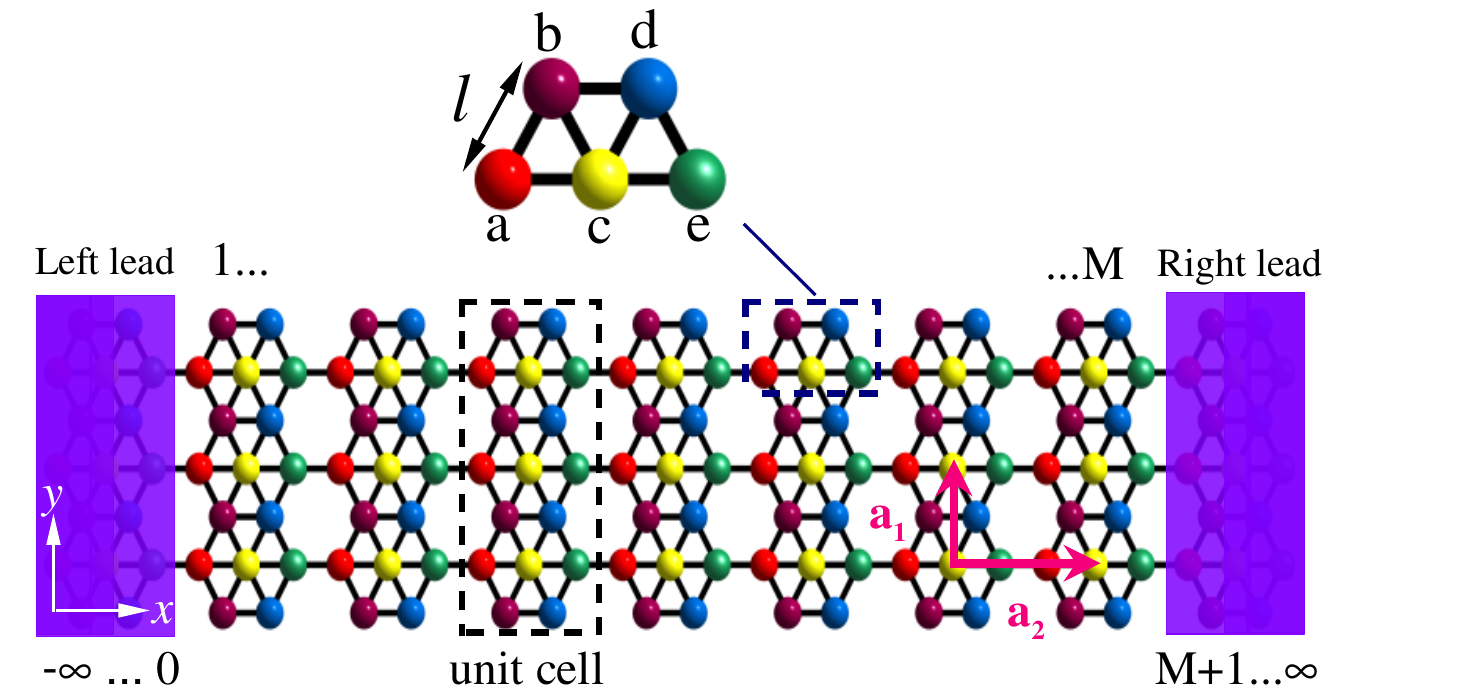}
\caption{\label{Fig1} Schematic illustration of an armchair BNR. The primitive cell indicated by dashed blue rectangle contains five atoms with labels b, d, a, c and e which are colored by purple, blue, red, yellow and green, respectively. Here, $\textbf{a}_1=\sqrt{3}\ell{\hat{y}}$ and $\textbf{a}_2=3\ell{\hat{x}}$ are the lattice basic vectors and $\ell\simeq{1.69}$ {\AA}  is the boron-boron atom distance. The middle part includes $M$ unit cells extended along the longitudinal direction so that each unit cell has N boron atoms.}
\end{figure}

The band structure of the monolayer $\beta _{12}$-borophene at high symmetry points including $K$ and $K'$ are described by a five-band tight-binding (TB) Hamiltonian that its hopping parameters have obtained by fitting first-principles calculation results \cite{bib29}. Although this Hamiltonian has only considered the $p_z$ orbitals of boron atoms, the recent first-principles calculation shows the electronic properties of $\beta _{12} $-borophene is well described by this model due to its perfect planar structure \cite{bib33}. Recently, three different models of TB Hamiltonian including homogeneous model, inversion symmetric (IS) and inversion non-symmetric (INS) models were introduced for $\beta _{12} $-borophene monolayer. However, the TB Hamiltonian based on INS model is experimentally closer to its real structure. Indeed, the inversion symmetry of $\beta _{12}$-borophene breaks due to effects of the Ag substrate and Dirac fermions become gapped \cite{bib29}. Considering this model, TB Hamiltonian of $\beta _{12}$-borophene in the presence of exchange and transverse electric fields is written as:
\begin{eqnarray}\label{Eq.1}
H =-{t}{\sum_{{\langle{i,j}\rangle},\alpha}}c^{\dagger}_{i\alpha}c_{j\alpha}+h\sum_{i\alpha}c^{\dagger}_{i\alpha}\sigma_{z}c_{i\alpha} \nonumber\\
+eE_{y} \sum_{i,\alpha}d_{i}c^{\dagger}_{i\alpha}c_{i\alpha}
+\sum_{i,\alpha}c^{\dagger}_{i\alpha}(\varepsilon_{i}+w_{i})c_{i\alpha},
\end{eqnarray}         
where the first term is the hopping energy between the nearest-neighbor (NN) sites which according to the primitive cell labelling of Fig.~\ref{Fig1}, are summarized as \cite{bib29}:\\[8pt]
$  t_{ab}=t_{de}=-2.04$ eV,  \:          $ t_{ac} = t_{ce} = -1.79$ eV, \:          \\[8pt]   	
$ t_{ae}= -2.12$ eV, \: \hfill(2) \\[8pt] 
$t_{bc} = t_{cd} = -1.84$ eV,  \:\: $ t_{bd} = -1.91$ eV. \\[8pt]       
Also, the operator $c^{\dagger} _{i\alpha}(c_{i\alpha})$ creates (annihilates) an electron at site $i$ with spin $\alpha$ and $\langle{i,j}\rangle$ denotes the sum over NN sites. The second term including the $z$ component of Pauli matrix $\sigma_z$ is due to the exchange magnetic field with strength $h$ which is induced through proximity to a ferromagnetic material \cite{bib34}. The third term accounts for the transverse electric field $E_y$ applied along the $y$ direction and $d_i$ is defined as the distance between site $i$ and the chosen origin. The last term includes the onsite energies for primitive cell (see Fig.~\ref{Fig1}) which based on inversion nonsymmetric model are given by \cite{bib29}:\\[8pt]
$\varepsilon_{a}=\varepsilon_{d}=0.196$ eV, \:  $\varepsilon_{b}=\varepsilon_{e}=-0.058$ eV,\:  \\[8pt] 
$\varepsilon_{c}=-0.845$ eV. \: \hfill(3) \\[8pt]
In the last term of Eq.~\ref{Eq.1}, $w_{i}$ is the on-site energy due to Anderson (short range) disorder which is distributed uniformly in the range $ [-w/2, +w/2]$ so that $w$ is the disorder strength.  
 
The setup of our system consists of an armchair BNR connecting to two semi-infinite leads as shown in Fig.~\ref{Fig1}. Such a system can be considered as a linear chain of repeated unit cells so that each of them consists of a zigzag array of N boron atoms [e.g., the unit cell specified in Fig.~\ref{Fig1} has N = 17 atoms]. There is a periodicity along with the longitudinal direction $\hat{x}$ that leads to $[H, p_x]=0$, so one can write the $k_x$-dependent Hamiltonian with the use of Bloch's theorem as:
\begin{equation}\label{sec2Eq.4}\tag{4}
H(k_{x})=H_{\ell,\ell+1}{e^{ik_{x}a}}+H_{\ell,\ell}+H_{\ell,\ell+1}^{\dagger}e^{-ik_{x}a},
\end{equation}
where $H_{\ell,\ell+1}$ and $ H_{\ell,\ell}$ describe respectively the interaction of each unit cell with the next one and interior interactions of the lth unit cell of $\beta_{12}$-borophene monolayer and here $a=0.29$ nm is the lattice constant. The energy dispersion of this structure can be easily calculated with diagonalizing above Hamiltonian.  
\subsection{Spin-dependent quantum conductance}
 For semi-infinite BNR with periodicity along one direction, there is an iterative method with very fast convergence which was first established by Sancho \textit{et al.} to obtain the quantum conductance in semi-infinite crystals \cite{bib35}. Using this method, one can easily calculate the surface Green’s functions of the left and right leads with the transfer matrix approach. Here, we focus on ballistic regime so that no inelastic scattering occurs along the length of the channel region and also phonon interactions under the assumption of low temperature is negligible.

Using first-principles calculations, for a carrier density of $n=3.4\times10^{13}$cm$^{-2}$, the intrinsic mobility of pristine $\beta _{12}$-borophene has been estimated about $\mu\approx540\,$cm$^{2}$V$^{-1}$s$^{-1}$ at room temperature \cite{bib36}. Thus, we can obtain a mean free path of about $\lambda=(h/2e){\mu}\sqrt{n/\pi}\approx36.74\,$nm \cite{bib37}. Considering a few unit cells, the channel length will be much smaller than $\lambda$, so that the required condition for ballistic regime is satisfied. In this case, we can apply the Landauer-B\"uttiker formula for the spin-dependent conductance as \cite{bib38}:
\begin{equation}\tag{5}
G^{s}(E)=\frac{e^{2}}{h}Tr[\Gamma_{L}^{s}(E)g^{s}(E)\Gamma_{R}^{s}(E)g^{s}(E))^{\dagger}],
\end{equation}
where $G^{s}$ indicates the electron conductance with spin $s$, $E$ is the electron energy, $g^{s}(g^{s\dagger})$ represent the retarded (advanced) Green’s function matrix and
\begin{equation}\tag{6}
 \Gamma^{s}_{L(R)}(E)=i\left({\Sigma^{s}_{L(R)}-(\Sigma^{s}_{L(R)})^{\dagger}}\right)
\end{equation}
is the broadening matrix due to strong couplings between the central region and left (right) lead of the system. Here, one can numerically calculate the spin-dependent self-energy $\Sigma^{s}_{L(R)}$ for left (right) lead via an iterative scheme \cite{bib39}. The self-energy at the contact point is given by:
\begin{equation}\tag{7}
\Sigma_{L}^{s}=H_{01}^{\dagger}g_{00}^{s}H_{01},
\end{equation}
where $g_{00}^{s}$ is computed via the following coupled equations:\\[8pt]
$ (E^{+}-H_{00})g_{00}^{s}=\textbf{I}+H_{-10}^{\dagger}g_{-10} ^{s},$\\[8pt]    
$ (E^{+}-H_{00})g_{-10}^{s}=H_{-10}^{\dagger}g_{-20}^{s}+H_{-10}g_{00}^{s},$\hfill(8)\\[8pt]
$\cdots$,\\[8pt]
$ (E^{+}-H_{00})g_{-\ell0}^{s}=H_{-10}^{\dagger}g_{-\ell-1,0}^{s}+H_{-10}g_{-\ell+1,0}^{s}\,,$\\[8pt]
where $\textbf{I}$  is a unit matrix and $E^{+}=E+i\eta$ so that $\eta $ is an infinitesimal real constant. There is a delicate approach to tackle the problem of a large number of iterative cells in semi-infinite terminals. According to this approach, one can express Green's
function of each unit cell in terms of the previous or next cell i.e., $g^{s}_{00}={\Lambda}g_{-10}^{s}$ or $ g^{s}_{-10}=\tilde{\Lambda} g^{s}_{00}$. Here, $\Lambda $ and $ \tilde{\Lambda} $ as transfer matrices are calculated via the following iterating procedure: \\[8pt]
$ \Lambda=t_{0}+{t}_{0}\tilde{t}_{1}+{t}_{0}{t}_{1}\tilde{t}_{2}+\ldots+{t}_{0}{t}_{1}{t}_{2} \ldots \tilde{t}_{n}, $\hfill(9)\\[8pt]
$ \tilde{\Lambda}=\tilde{t}_{0}+\tilde{t}_{0}{t}_{1}+\tilde{t}_{0}\tilde{t}_{1}{t}_{2}+\ldots+\tilde{t}_{0}\tilde{t}_{1}\tilde{t}_{2}\ldots {t}_{n}, $\hfill(10)\\[8pt]
where $ t_{i} $ and $\tilde{ t_{i}}$ comes out of following recursion equations: \\[8pt]
$ t_{i}=(I-t_{i-1}\tilde{t}_{i-1}-\tilde{t}_{i-1}t_{i-1})^{-1}t_{i-1}^{2}, $ \hfill(11)\\[8pt]
$\tilde{t_{i}}=(I-t_{i-1}\tilde{t}_{i-1}-\tilde{t}_{i-1}t_{i-1})^{-1}{\tilde{t}_{i-1}^{2}}, $ \hfill(12)\\[8pt]
and\\[8pt]
$ t_{0}=(E^{+}I-H_{00})^{-1}H_{-10}^{\dagger},$\hfill(13)\\[8pt]
$\tilde{t_{0}}=(E^{+}I-H_{00})^{-1}H_{-10}.$\hfill(14)\\[8pt]
The process has to repeat until $ t_{n}, \tilde{t}_{n} \leq \delta $, where $ \delta $ is a small arbitrary constant. Based on transfer-matrix approach, the surface (zeroth-cell) Green's function can be extracted by:\\[8pt]
$g^{s}_{00}(E)=[E^{+}\textbf{I}-H_{00}-H^{\dagger}_{-10}\tilde{\Lambda}]^{-1}\,. $\hfill(15)\\[8pt]
Besides, the self-energy due to the scattering into the right lead can be obtained in a similar way.
The surface Green's function is able to obtain step by step for any arbitrary unit cell inside the transport channel by translating $ g_{00}^{s}(E) $ as follows \cite{bib40}:
\begin{equation}\tag{16}
g^{s}_{\ell+1,0}(E)=[{E^{+}}\textbf{I}-H_{\ell,\ell}-H_{\ell,\ell+1}^{\dagger}g^{s}_{\ell,0}H_{\ell,\ell+1}]^{-1}\,.
\end{equation}
%$ g^{s}_{\ell,\ell}(E_{F})=[(E_{F}+i\eta)I-H_{\ell,\ell}-H_{\ell,\ell+1}g^{R}_{\ell+1,\ell+1}H_{\ell,%\ell+1}^{\dagger}]^{-1}\,.$ \\[4pt]
%\hfill(16)\\[8pt]
Density of states (DOS) for a semi-infinite crystal's layer in terms of the Green’s function, is given by well-known formula:
\begin{equation}\tag{17}
\rho^{s}(E)=-\frac{1}{\pi}Im[Tr(g^{s}(E))].
\end{equation}
In this case, the spin-dependent local density of states (LDOS) at site $n$ is defined as
\begin{equation}\tag{18}
\rho_{n}^{s}(E)=-\frac{1}{\pi}Im[g_{(n, n)}^{s}(E)].
\end{equation}
The spin-dependent local current density at energy $E$ between two neighboring sites $m$ and $n$ can be written as \cite{bib41}
\begin{equation}\tag{19}
J_{m\rightarrow{n}}^{s}(E)=\frac{4e}{\hbar}Im[H_{mn}G_{nm}^{c, s}],
\end{equation}
where $G^{c, s}={g^s}{\Gamma^{s}_{L}}{g^{s\dagger}}$ is the spin-dependent correlation function and $H_{mn}$ is the corresponding matrix element of the Hamiltonian. Finally, the spin polarization in terms of spin-dependent conductance is calculated as \cite{bib42, bib43}
 \begin{equation}\tag{20}
P_{s}(E)=\frac{G^{\uparrow}-G^{\downarrow}}{G^{\uparrow}+G^{\downarrow}}. 
\end{equation}
Note that $-1\leq{P_{s}}\leq1$ and $P_{s}>0(<0)$ shows the dominant spin polarization is
the spin-up (down) state.                             
\section{RESULTS AND DISCUSSION}\label{sec3}
\begin{figure}[t]
\centering
\includegraphics[scale=0.3]{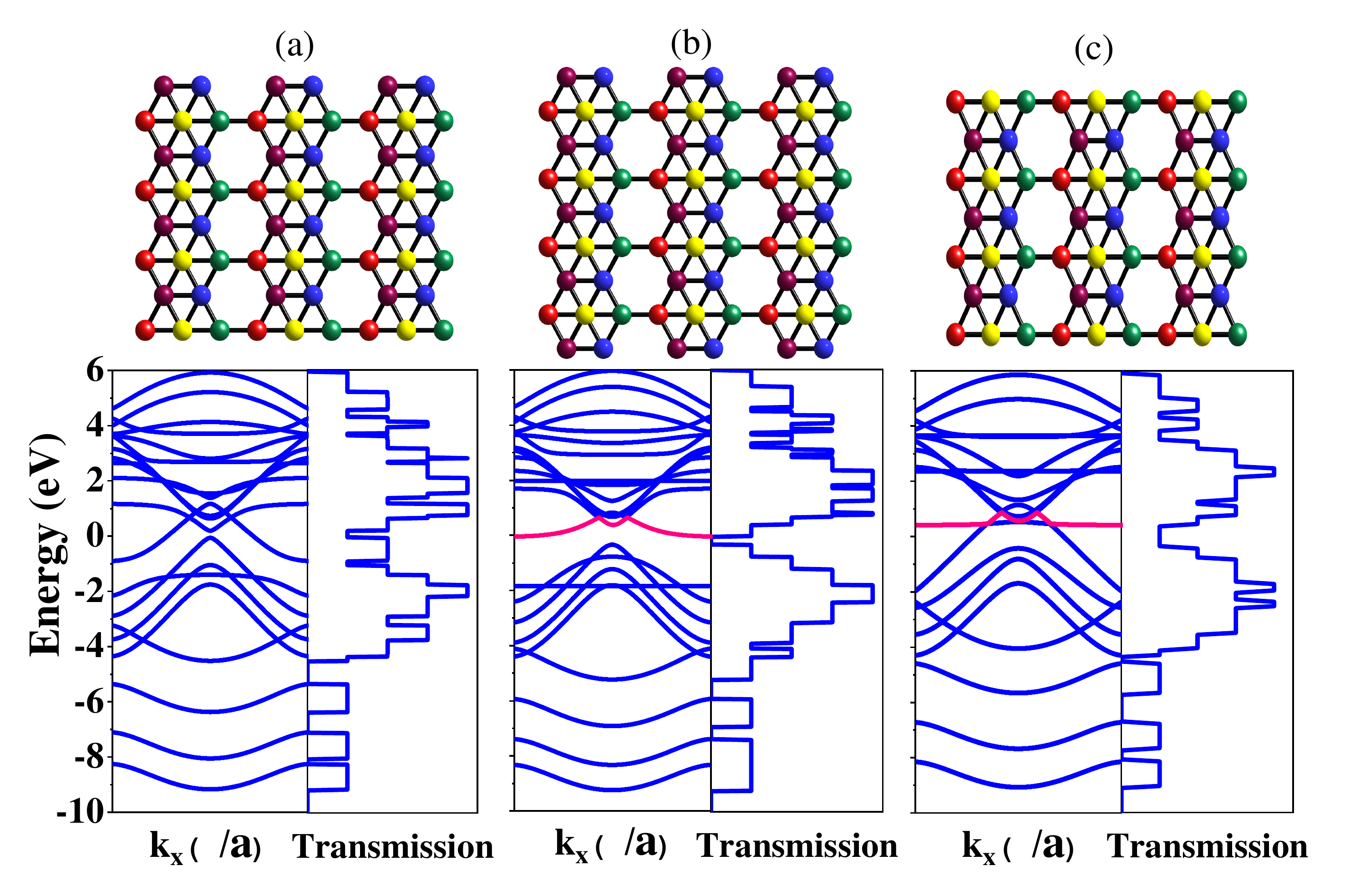}
\caption{\label{Fig2}Top view of the atomic configurations (upper) and the band structure and transmission channel (lower) for ABNRs with (a) AB (b) AA and (c) BB edges. There are a flat band near the Fermi (zero) energy for AA edges and a quasi flat band for BB edges indicated by pink color.}
\end{figure}

We now present our results calculated by employing TB Hamiltonian and NEGF method as stated before. In this research, we only focus on the armchair BNRs (ABNRs) because they are more interesting than zigzag BNRs (ZBNRs) due to the tunable bandgap \cite{bib29}. The electronic properties of 2D hexagonal materials greatly depend on their edge structure. Experimentally, BNRs is produced by cutting a borophene sheet that each of them have different edge shapes with distinct electronic properties, so the study of edge-dependent electronic properties is crucial. In order to clarify the influence of different edges in ABNRs, the band structure and transmission channel for three infinite ABNRs with different widths or edges have been illustrated in Fig.~\ref{Fig2}. As seen in this figure, the edges are terminated by armchair and flat shapes which hereafter are denoted by A and B, respectively. It is observed in Fig.~\ref{Fig2} that the edges of ribbon with the width $N=22$ atoms are AA and the ones with widths $N=20$ and $N=18$ are AB and BB, respectively. 

Figures \ref{Fig2}(a) and \ref{Fig2}(c) show that ABNRs with AB and BB edges are metal while ABNRs with AA edges in Fig.~\ref{Fig2}(b) are semiconductor corresponding to their ribbon width [for widths with $N<27$ atoms]. Also, our calculations show that ZBNRs are metals similar to zigzag graphene nanoribbons (ZGNRs) and emerge three-band touching points or triplet fermions in their band structure. However, the degeneracy of three-band touching is lifted for ABNRs so that the bands shown in Fig.~\ref{Fig2} play a crucial role in their electronic structure. For widths with $N>27$ atoms, all the ABNRs are metals and for this case cutting a $\beta_{12}$ sheet in different directions cannot induce a band gap. This result is in good agreement with the recent experimental reports \cite{bib26} and first-principle calculations \cite{bib44}. Here, the low-lying energy bands are almost flat at $k=\pm\pi/a$ so that the group velocity $v_g=\hbar^{-1}{\partial{E}/\partial{k}}$ is vanishingly small and it gives rise to the localization and accumulation of electrons at the edge of the ribbons, which was confirmed experimentally with the help of the scanning tunneling microscopy (STM) results in the boundaries a $\beta_{12}$ sheet \cite{bib4, bib26}. Meanwhile, the band dispersions for the crossing points seen in Fig.~\ref{Fig2} are similar to the massless Dirac fermions that are no longer flat in compared with ZGNRs. As shown in Figs, 2(b) and 2(c), near the Fermi energy ($E$ = 0), there are a flat and a quasi-flat bands for nanoribbons with AA and BB edges, respectively. These bands may reveal the novel quantum and topological phases similar to topological kagome magnet \cite{bib45}. Indeed, recent studies show that the flat band with non-zero group velocity and the massive Dirac fermion near Fermi energy originates from the band with intrinsic Berry curvature \cite{bib46}.    
\begin{figure}[t]
\centering
\includegraphics[scale=0.43]{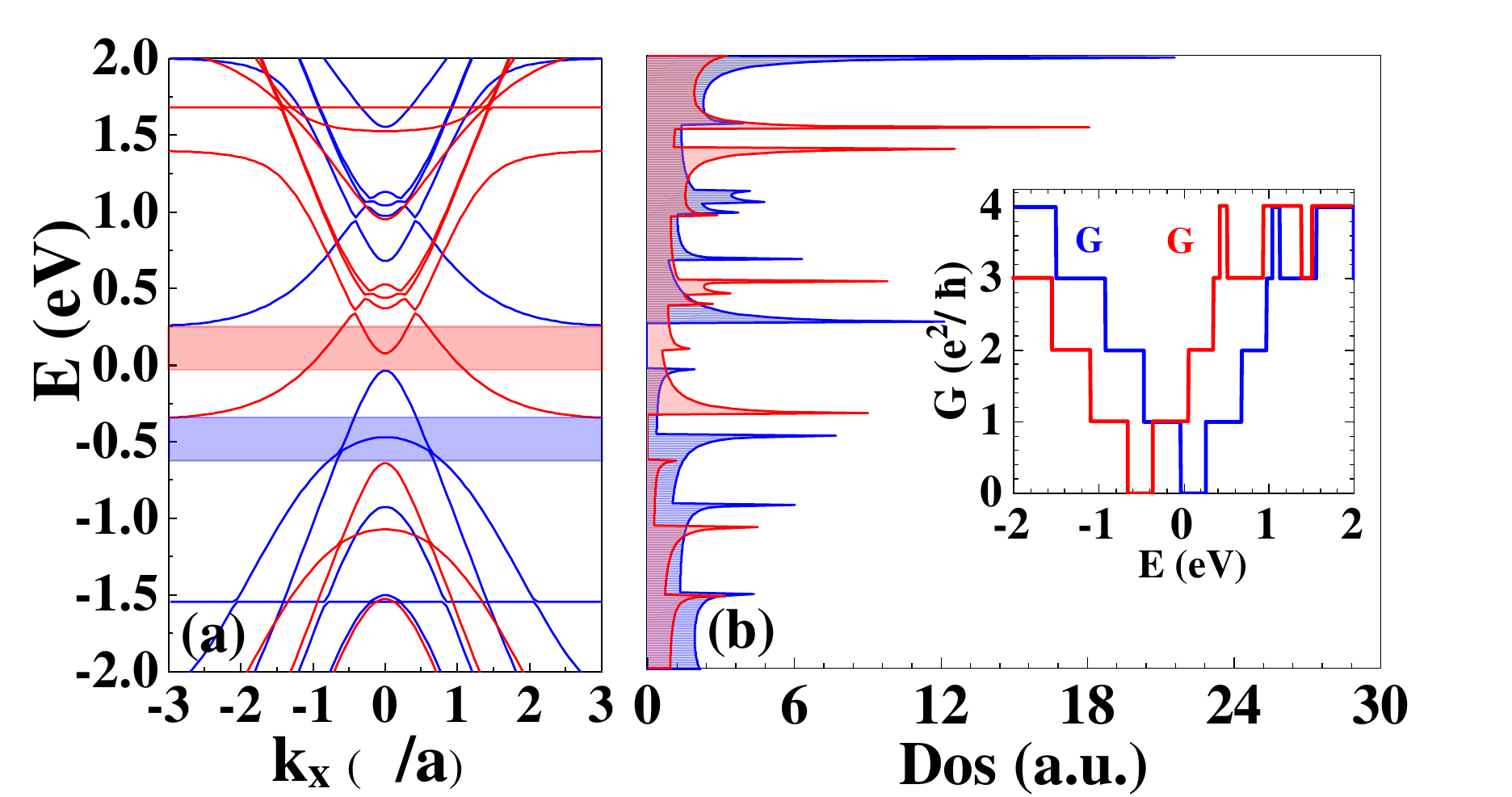}
\caption{\label{Fig3}(a) Band structure and (b) DOS for a 22-ABNR with AA edges when a nonlocal exchange field of strength $h=0.3$ eV is applied. The fully polarized regions in terms of spin-up and spin-down are shown in blue and red, respectively. It is obvious that the Van Hove peaks (singularities) of DOS coincide with the energy subbands. The inset of (c) indicates the spin-dependent conductance $G^{\uparrow}$ and $G^{\downarrow}$ as a function of the energy of the incoming electrons.}
\end{figure}   
It is worth mentioning that ABNRs can be the host of nontrivial edge states that are protected by time-reversal symmetry (TRS) and play an important role in the transport properties of system \cite{bib47}. In order to break TRS or spin inversion symmetry, an exchange field can be induced  via the proximity effect produced by a ferromagnetic insulator substrate. In contrast with other techniques for breaking TRS such as transition metal atoms doping \cite{bib48, bib49} and atomic bulking \cite{bib50, bib51}, this approach is an external manipulation with short-range interaction which has no effect on the mobility of electrons from the perspective of electronic applications. 

Taking into account only a nonlocal exchange field $h=0.3$ eV throughout a 22-ABNR with AA edges and TRS breaking, the spin-up and spin-down states are split and shifted toward opposite directions on the $E$ axis as shown in Fig.~\ref{Fig3}(b). Here the nonlocal exchange field means that the exchange field is applied to the whole system (i.e., the channel and both leads) and the local exchange field means that the exchange field is applied only to the channel or some parts of it. The induced spin polarization due to this asymmetry is reflected in the density of states as the spin polarized Van Hove peaks (singularities) which can be clearly observed in Fig.~\ref{Fig3}(c). These peaks were previously seen in one dimensional (1D) systems composed of graphene and carbon nanotubes \cite{bib52, bib53}. One can prove from this figure that $E(s,k_{x})=E_{-s,-k_{x}}-2sh$ so that the induced spin gap is proportional to the strength of the exchange field $E^{s}_g\propto{h}$ and also the Bloch periodic boundary condition leads to the invariant band structure under translation $2\pi/a$. Here, the direction of spin polarization depends on the energy of incoming electrons, which can be tuned by an external gate voltage in practice. 
\begin{figure}[t]
\centering
\includegraphics[scale=0.75]{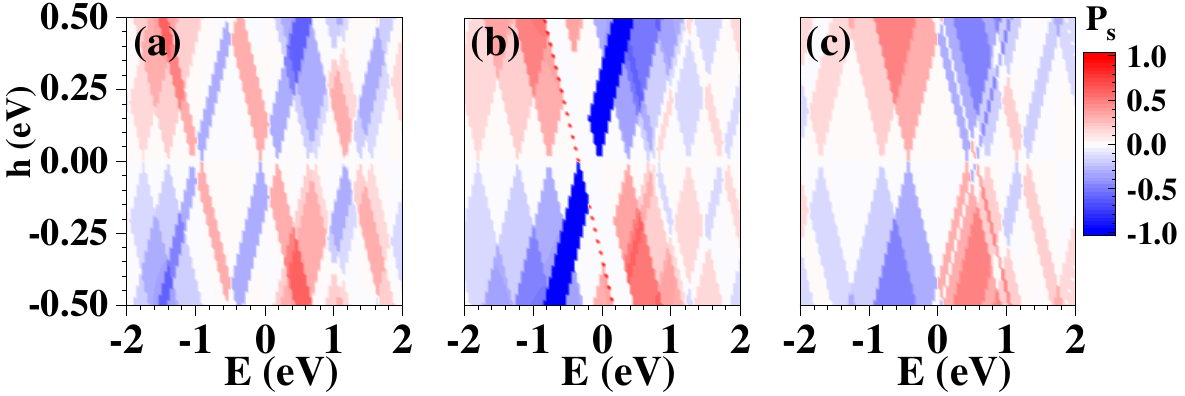}
\caption{\label{Fig4} Contour plot of spin polarization $P_{s}$ as a function of the strength of exchange field $h$ and the energy of incoming electrons $E$ for the different edge shapes including (a) 20-ABNR with AB edges (b) 22-ABNR with AA edges (c) 18-ABNR with BB edges.}
\end{figure}
As can be observed in the inset of Fig.~\ref{Fig3}(c), the conductance between 0.2 eV $<E<$ 0.4 eV (-0.4 eV $<E<$ -0.2 eV) is completely polarized so that only the electrons with spin-down (spin-up) can pass through the 22-ABNRs in this energy region of transmitted electrons. This result shows that ABNRs can act as a controllable half-metal system because they have a metallic state for one spin direction and an insulating state for the opposite spin direction at the same time \cite{bib54}. 

In order to investigate the edge-dependent spin polarization of ABNRs, the contour plot of spin polarization $P_s$ for the different edge shapes mentioned in Fig.~\ref{Fig2} as function of the strength of exchange field $h$ and the energy of incoming electrons is illustrated in Fig.~\ref{Fig4}. As observed, the perfect spin polarization is achievable for the various edge shapes of ABNRs so that the direction of spin polarization is variable for the different energies of incoming electrons.  Notice that $E$ can be experimentally tuned by using a back gate voltage, so the ABNRs can act as a controllable spin filter for spintronic applications.

In order to control the spin transport properties, the induction of half-metallicity in 2D materials is more important for spintronics. In 2006, researchers suggested that the half-metallicity can be induced in ZGNRs when a transverse electric field is applied \cite{bib55}. To assess this effect, a transverse electric field along the $y$-direction $\textbf{E}=(0,E_{y},0)$ is applied to the middle part (channel) of 22-ABNRs in combination with a nonlocal exchange field.  
\begin{figure}[ht]
\centering
\includegraphics[scale=0.3]{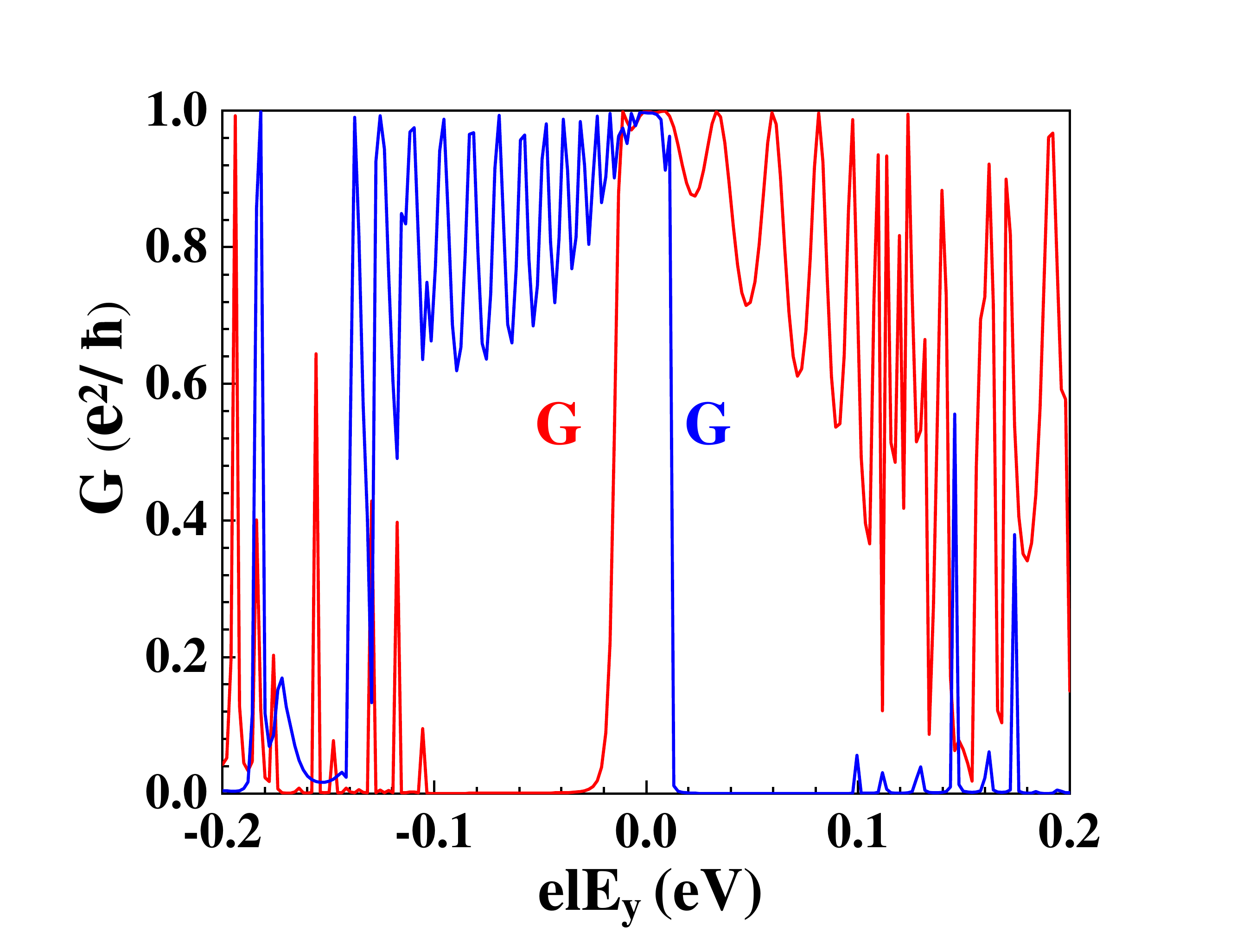}
\caption{\label{Fig5} Spin-dependent conductance for a 22-ABNR as a function of normalized transverse electric field $e{\ell}{E_y}$ in the presence of a nonlocal exchange field $h=0.2$ eV at a constant energy $E=-0.2$ eV of incoming electrons.}
\end{figure}

Figure \ref{Fig5} shows the spin-dependent conductance as a function of the transverse electric field $E_y$ in the presence of a nonlocal exchange field $h=0.2$ eV for a constant energy of incoming electrons (i.e, $E=-0.2$ eV). As seen, the electrons with spin down can only pass through ABNRs for the electric field in $+\hat{y}$ direction ($0.02$ eV $<e{\ell}{E_y}<0.1$ eV) so that the system can act as a perfect filter or spin polarizer for spin-down. Reversing the electric field direction (i.e, $-\hat{y}$) for $-0.1$ eV $<e{\ell}{E_y}<-0.02$ eV, the spin direction of transmitted electrons through the ABNRs can be easily changed from down to up. This feature is noteworthy because it leads to the electric control of perfect spin polarization which is an empirically desirable effect for spintronic nanodevices. Also, this result shows the half-metallicity can be triggered in ABNRs by a transverse electric field.  
 
\begin{figure}[t]
\centering
\includegraphics[scale=0.75]{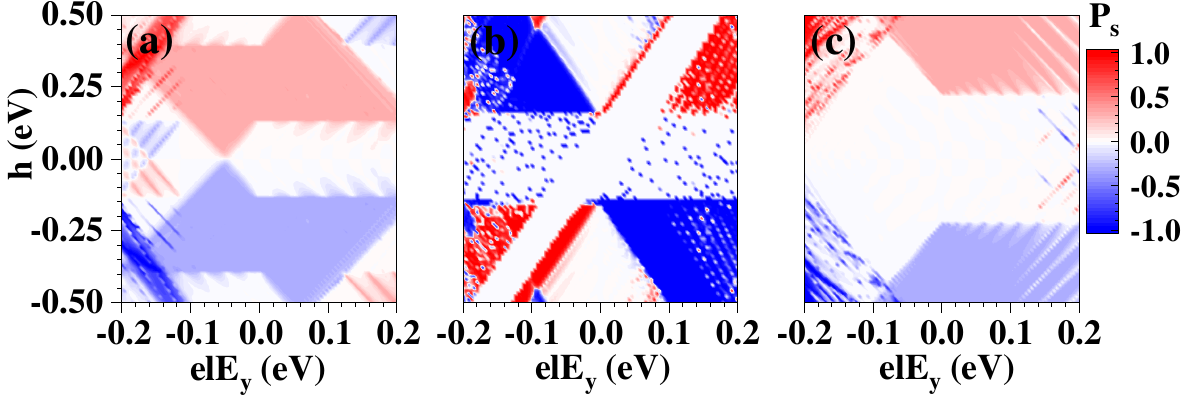}
\caption{\label{Fig6} Contour plot of spin polarization $P_{s}$ as functions of the strengths of transverse electric field $e{\ell}E_y$ and the exchange magnetic field at a constant energy $E=-0.2$ eV of incoming electrons for the different edge shapes of ABNRs with (a) AB (b) AA (c) BB edges.}
\end{figure}

In order to clarify the simultaneous influence of electric and exchange fields on the spin transport in ABNRs, the contour map of spin polarization $P_{s}$  for the different edge shapes of ABNRs as a function of $e{\ell}{E_y}$ and $h$ at a fixed energy of incoming electrons (i.e, $E=-0.2$ eV) is shown in Fig.~\ref{Fig6}. As observed, the perfect spin polarization for the different spin directions in ABNRs with various edge shapes is attainable by adjusting the strength and the orientation of transverse electric and exchange fields.                        

We have so far studied the effect of nonlocal exchange field on the spin transport of ABNRs with different edge shapes. However, the effect of local exchange field that breaks locally TRS only on the edges of ABNRs may lead to promising applications in spintronics. From the technological and experimental aspects, there are two important reasons for studying this effect in 1D structures exhibiting edge states. First, the existence of substrate and gates may not always provide access to the entire sample and second the edge manipulations of a large system cannot destroy its topological properties. 

\begin{figure}[t]
\centering
\includegraphics[scale=0.8]{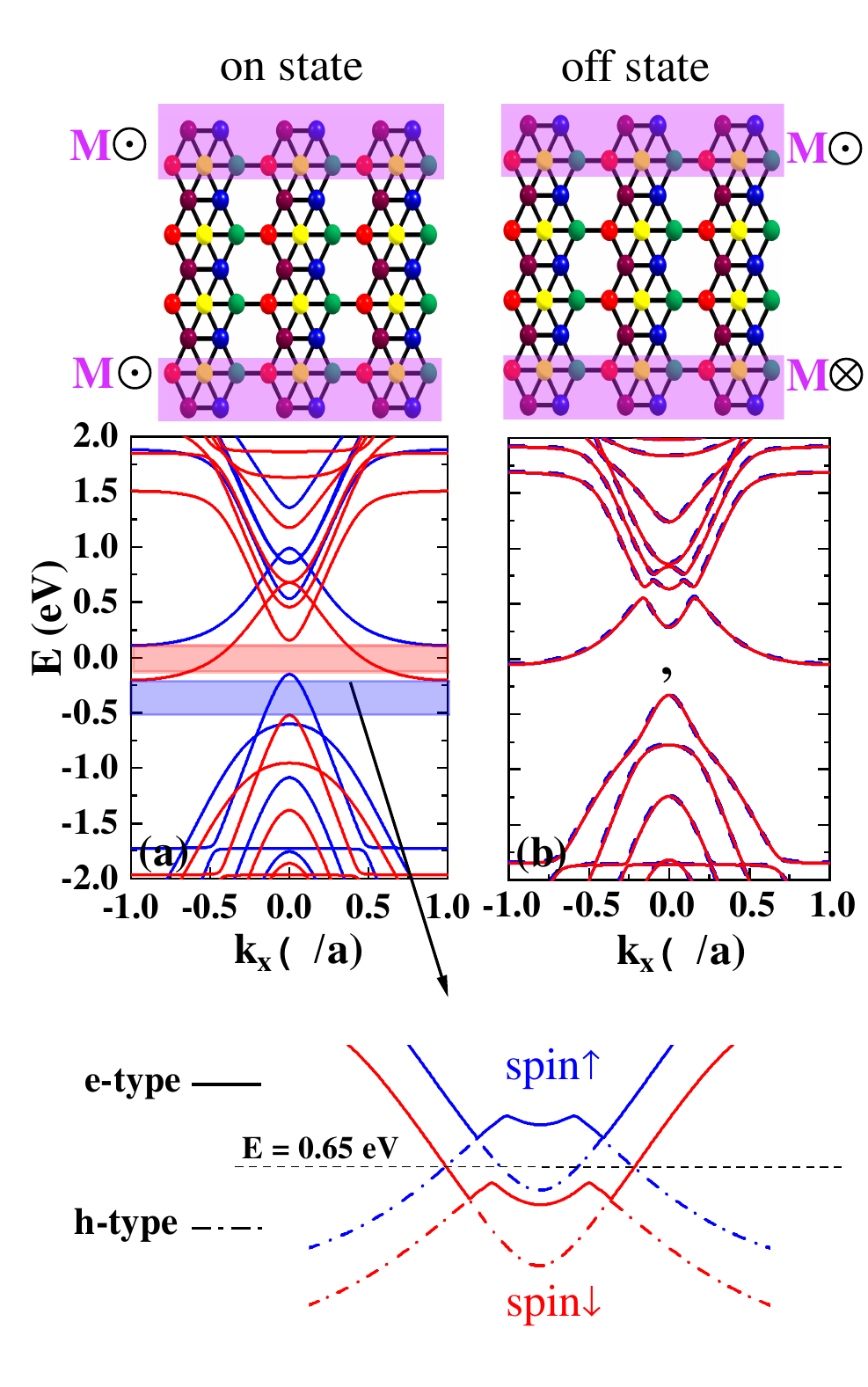}
\caption{\label{Fig7} Band structure for a 22-ABNR in the presence of local exchange fields on both edges of nanoribbon with the strength of $h=0.3$ eV for parallel (left panel) and antiparallel (right panel) configurations. The local exchange fields are induced by ferromagnetic stripes which are shown by magenta color. There are the gapless edge states for parallel configuration while they disappear for antiparallel configuration. Here, the transverse electric field is zero (i.e, $E_y=0$) and the magnified area (lower panel) indicates the band structure about the $E=0.65$ eV axis.}                     
\end{figure}

To investigate the influence of local exchange fields, two ferromagnetic strips are deposited on the both edges of 22-ABNR in the absence of electric field (i.e, $E_y=0$) with parallel and anti-parallel configurations as shown in Fig.~\ref{Fig7}. The orientation of exchange fields on both edges of ABNR are the same and opposite for parallel and anti-parallel configurations, respectively. As seen in  the left bottom panel, the spin-up and spin-down bands for parallel configuration are split and shifted toward opposite directions on the $E$ axis similar to nonlocal exchange field [see Fig.~\ref{Fig3} for comparison]. In other words, the edge states for the spin-up and spin-down contribute to a spin current, see the edge states specified in this figure which corresponds to the \q{on state}. Here, the lowest electron-type (positive effective mass) and hole-type (negative effective mass) bands for the different spin directions are illustrated in the magnified area of Fig.~\ref{Fig7} which intersect each other at $k=\pm{\pi/5a}$. This shows that the spin filtering occurs for spin-up and spin-down with a minor symmetry about the $E=0.65$ eV axis so that for $E>0.65$ eV ($E<0.65$ eV) the electrons can only pass with spin-up (spin-down) through the nanoribbon. 
For anti-parallel configuration as seen in the right bottom panel of Fig.~\ref{Fig7}, the bands for spin-up and spin-down are degenerate. This corresponds to the \q{off state} because there is a small gap between the edge modes that leads to a zero spin polarized current in this energy range of incoming electrons. Notice that the magnitude of induced gap increases when the regions of the applied exchange fields on both edges become larger. This means that the penetration length of edge modes at armchair edges is long. In addition, the above results show that ABNRs can act as a giant magnetoresistance because the spin conductance is tuned or induced by changing the direction of exchange fields on both edges of nanoribbon. 

\begin{figure}[t]
\centering
\includegraphics[scale=0.3]{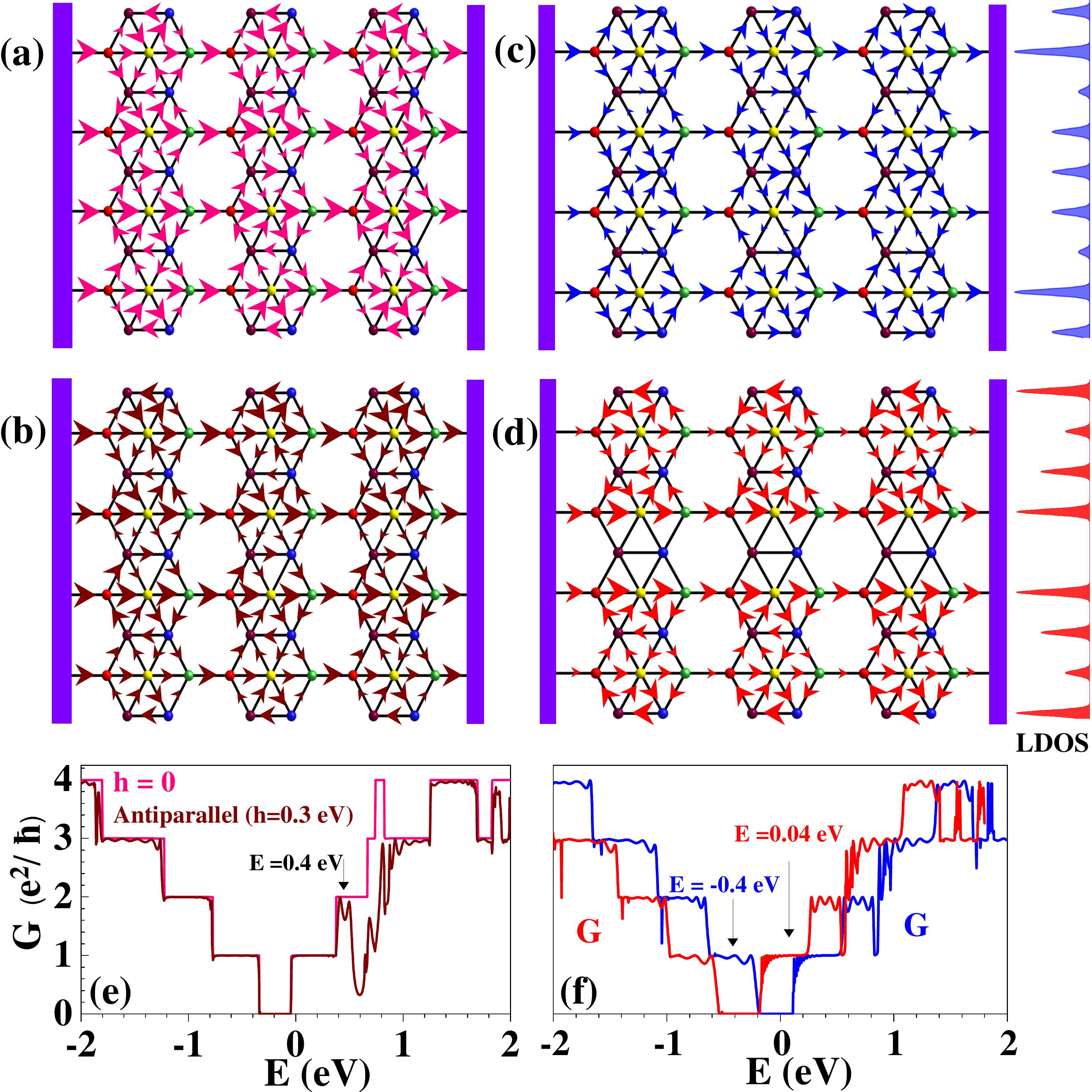}
\caption{\label{Fig8} Local current distribution (a) in the absence (i.e., $h=0$) and (b) presence of local exchange fields (i.e., $h=0.3$ eV) with antiparallel configuration for incoming electron energy $E=0.4$ eV. For parallel configuration in which spin polarization can occur (see Fig.~\ref{Fig7}), spin-up and spin-down local current distributions are shown in (c) and (d), respectively. Here, $h=0.3$ eV and the incoming energies are chosen as $E=-0.4$ eV for spin-up and $E=0.04$ eV for spin-down. The values of chosen incoming energy for $h=0$, parallel and antiparallel configurations, are depicted by black arrows in conductance graphs shown in (e) and (f). Note that $G^{\uparrow}$ and $G^{\downarrow}$ for $h=0$ and antiparallel configuration are completely coincident. For parallel configuration at $E=-0.4$ eV ($E=0.04$ eV) only electrons with spin-up (spin-down) can pass through the ABNR.}                     
\end{figure}

In Fig.~\ref{Fig8}, we study the spatial distribution of local current in the presence and absence of exchange field. Here, the magnitude of the local current between two neighboring sites is depicted by the size of arrow. It is observed that, unlike GNRs, the existence of \q{c} atom in the center of honeycomb lattice of BNRs (see Fig.~\ref{Fig1}) leads to paths include current loops. As expected and shown in Fig. 8(a), there are current paths along the transport direction which are equivalent to each other in the absence of exchange fields. In this case, the local spin-up and spin-down current distributions are the same because the spin-up and spin-down conductances are totally coincident (see, Fig.~\ref{Fig8}(e)). As can be clearly seen in Fig.~\ref{Fig8}(b), the currents in the presence of exchange fields (i.e., $h=0.3$ eV) with antiparallel configuration are nearly the same with $h=0$. The conductance dip appeared in Fig.~\ref{Fig8}(e) for this configuration relative to $h=0$ is due to the channel mismatch with leads. For parallel configuration, the local spin-up and spin-down current distributions are shown in Figs.~\ref{Fig8}(c) and \ref{Fig8}(d), respectively. Here, we choose the values of incoming electron energy (i.e., $E = -0.4$ and 0.04 eV) such that the electron currents become fully spin polarized with spin-up or spin-down (see, Fig.~\ref{Fig8}(f)). In this configuration, current paths are not equivalent and the local spin-up (spin-down) currents at the edges of ABNRs are bigger (smaller) than other paths. This fact is well followed by local density of states (LDOS) shown in the right side of Figs.~\ref{Fig8}(c) and \ref{Fig8}(d). As shown in Fig. 8(d), for spin-down, the current paths are not formed in the central region of ABNRs. In other words, these paths are not effective for spin-down transport in ABNRs.          

\begin{figure}[t]
\centering
\includegraphics[scale=0.55]{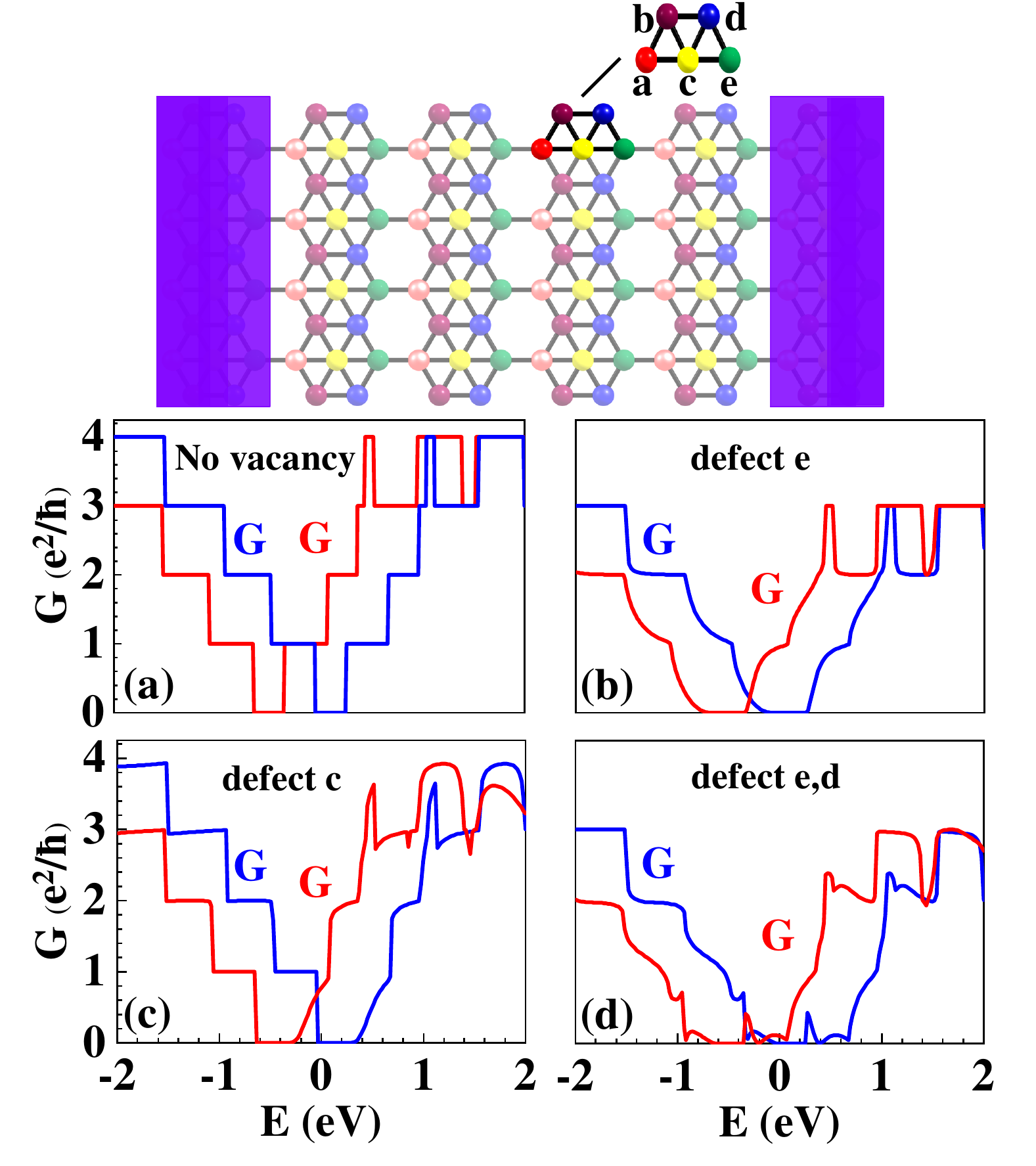}
\caption{\label{Fig9} Spin-dependent conductances $G^{\uparrow}$ and $G^{\downarrow}$ of 22-ABNR (a) without vacancy (b) with \q{e} atom vacancy (c) with \q{c} atom vacancy (d) with \q{e} and \q{d} atoms vacancy as a function of the energy of the incoming electrons in the presence of a nonlocal exchange field of strength $h=0.3$ eV.}
\end{figure}

Borophene fabrication can be associated with some defects which were experimentally observed by STM images \cite{bib56}. Recently, the intrinsic defects of borophene were investigated at the atomic scale with ultrahigh vacuum STM and first principles calculations \cite{bib57}. They found that these defects can even lead to new phases in borophene. These results motivated us to study the effects of simple defects on the spin conductance of ABNR. Although it seems that a single defect (e.g., one-atom or two-atom vacancy) is not realistic, its study will pave the way to understand more complex defects. 

In Fig.~\ref{Fig9}, we plot the spin-dependent conductance of a 22-ABNR as a function of the energy of the incoming electrons in the presence of different defects. To better compare the impact of these defects, the spin-dependent conductance without any defects (i.e, perfect 22-ABNR) in the presence of a nonlocal exchange field of strength $h=0.3$ eV has been illustrated in Fig.~\ref{Fig9}(a). As the simplest defect, we examine the effect of a single \q{e} atom vacancy on the spin-dependent conductances $G^{\uparrow}$ and $G^{\downarrow}$ as shown in Fig.~\ref{Fig9}(b). As seen, spin transport gap for spin-up and spin-down increases which is due to quasi-localized states created by the vacancy \cite{bib58}. For the \q{c} atom vacancy as seen in Fig.~\ref{Fig9}(c), this defect has a slighter effect on the spin-dependent conductance than the \q{e} atom vacancy. Because the \q{e} atom vacancy breaks the inversion symmetry of atoms in the honeycomb lattice while the \q{c} atom vacancy preserves such symmetry. For the two-atom vacancy (i.e, \q{e} and \q{d} atoms vacancy) as observed in Fig.~\ref{Fig9}(d), the spin transport gap for spin-up and spin-down becomes larger than the \q{e} atom vacancy because two-atom vacancy creates more quasi-localized states in this system. In summary, the study of the impact of various defects on the spin-dependent conductance of ABNRs shows that there is still the perfect spin polarization in defective ABNRs. 
\begin{figure}[t]
\centering
\includegraphics[scale=0.5]{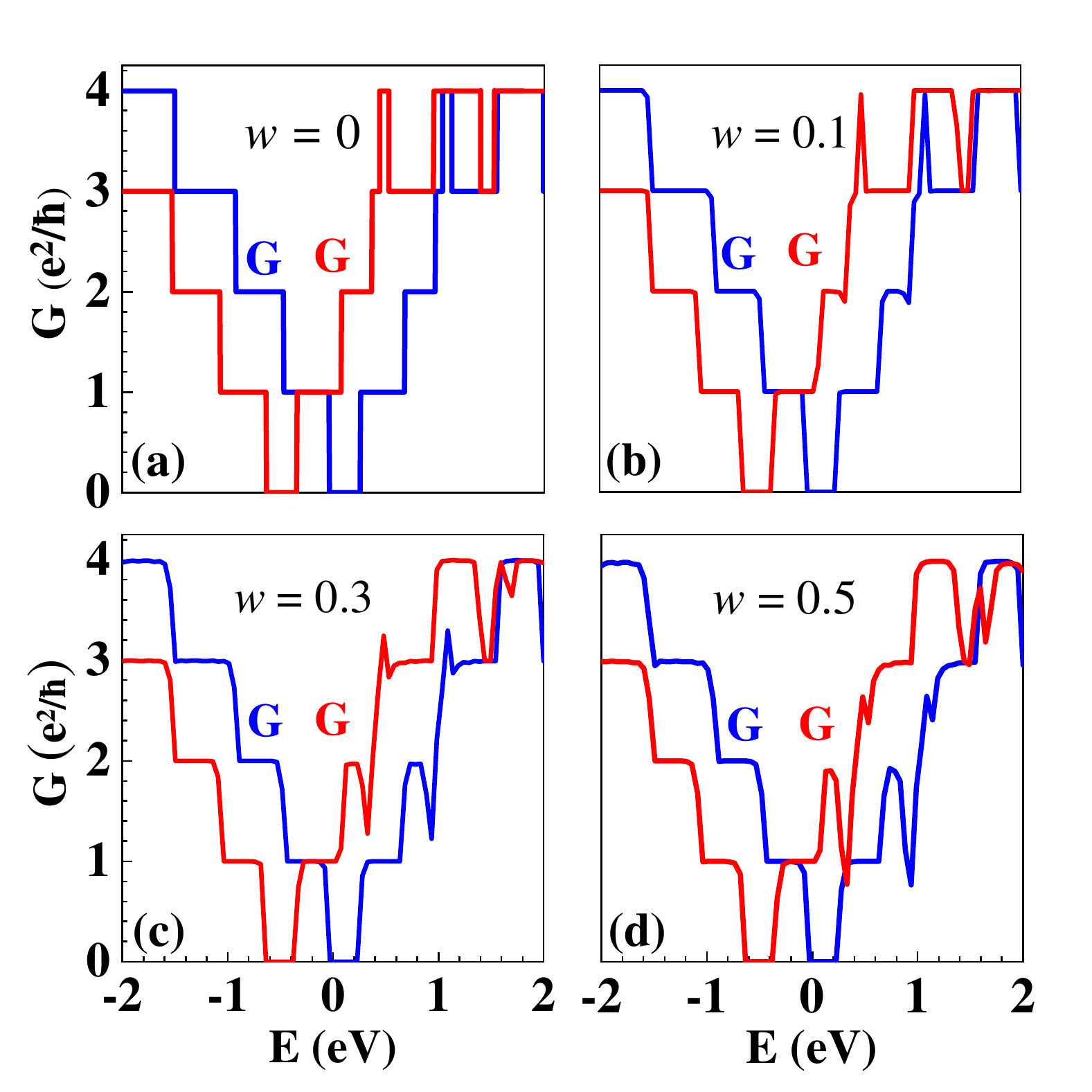}
\caption{\label{Fig10} Spin-dependent conductances $G^{\uparrow}$ and $G^{\downarrow}$ of 22-ABNRs as a function
of the energy of incoming electrons with a nonlocal exchange field of strength $h=0.3$ eV (a) in the absence of the Anderson disorder (i.e, $w=0$) and in the presence of an Anderson disorder of strength (b) $ w=0.1$ eV (b) $w=0.3$ eV (d) $ w=0.5$ eV.}  
\end{figure}

For real BNRs, there will always be some uncontrollable defects that are randomly distributed in the sample due to lattice distortion or impurity. For this purpose, we investigate the spin-dependent conductance of ABNRs under the influence of the different strengths of Anderson (short range) disorder as shown in Fig.~\ref{Fig10}. In Fig.~\ref{Fig10}(a), we plot the spin-dependent conductance of a 22-ABNR in the absence of the Anderson disorder (i.e, $w=0$ same as Fig.~\ref{Fig9}(a)) for comparison with other figures. In order to achieve an accurate result, the spin-conductance has been averaged over 1000 random disorder configurations. It is obvious that the spin-conductances $G^{\uparrow}$ and $G^{\downarrow}$ for the different strengths of disorder changes slightly. In fact, the obtained results show that the perfect spin polarization is robust against the different strengths of disorder and still ABNRs can act as a perfect spin filter for a certain energy of the incoming electrons.     
 
\section{SUMMARY}\label{sec4}

We have studied charge and spin transport in ABNRs with different edge shapes in the presence of transverse electric and exchange fields. When only a nonlocal exchange field is applied, ABNRs can act as a perfect spin filter for both spin-up and spin-down that the spin direction of transmitted electrons can be tuned by adjusting the energy of incoming electrons. In the simultaneous presence of a transverse electric field and a nonlocal exchange field, the spin filtering is also electrically controllable and the half-metallicity can be induced in ABNRs. We also investigated the effect of local exchange field by exposing the edges of ABNRs to ferromagnetic strips with parallel and antiparallel configurations. In this case, the spin-up and spin-down bands for parallel configuration are split and their corresponding edge states contribute to a spin current. However, there is a small band gap between the edge states for anti-parallel configuration which leads to a zero spin current. As a result, these edge manipulations emerge a giant magnetoresistance in ABNRs. Finally, we showed that the perfect spin polarization induced in ABNRs is robust against the edge vacancies and the different strengths of disorder which are randomly distributed in the nanoribbon.  

\bibliographystyle{apsrev4-2} %the RSC's .bst file
\bibliography{aps} %You need to replace "aps" on this line with the name of your .bib file

\end{document}